\documentclass[journal,twoside,web]{ieeeconf}
\usepackage[
top    = 0.75 in,
bottom = 0.75 in,
left   = 0.75 in,
right  = 0.75 in]{geometry}
\usepackage{amsmath,amssymb,amsfonts,graphicx,url,color}
\usepackage{cite}
\usepackage{mathtools}

\newtheorem{proposition}{Proposition}[section]
\newtheorem{problem}{Problem}[section]
\newtheorem{remark}{Remark}[section]
\usepackage{mathtools}
\graphicspath{{./},{./figures/}}

\newcommand{\ud}{d} 

\newcommand{\Area}{{\mathcal A}}

\newcommand{\op}[1]{\operatorname{#1}}

\newcommand{\trace}{\op{Tr}}

\newcommand{\mR}{\mathbb{R}}
\newcommand{\lam}{{\ell_r}}

\newcommand{\nablax}{{\partial_x}}
\newcommand{\nablay}{{\partial_y}}

\newcommand{\W}{{\rm W}}

\usepackage{graphicx}
\title{Geometry of finite-time thermodynamic cycles\\
with anisotropic thermal fluctuations}

\date{January 2022}

\begin{document}

\author{Olga Movilla Miangolarra$^{1}$, Amirhossein Taghvaei$^{2}$, Yongxin Chen$^{3}$, and Tryphon T. Georgiou$^{1}$
 \thanks{$^{1}$
Department of Mechanical and Aerospace Engineering, University of California, Irvine, CA 92697, USA; \{omovilla,tryphon\}@uci.edu}
\thanks{$^2$Aeronautics and Astronautics  Department, University of Washington, Seattle, WA 98195, USA; amirtag@uw.edu}
\thanks{$^3$
 School of Aerospace Engineering, Georgia Institute of Technology, Atlanta, GA 30332, USA; yongchen@gatech.edu}}

\maketitle

\begin{abstract}
In contrast to the classical concept of a Carnot engine that alternates contact between heat baths of different temperatures, naturally occurring processes usually harvest energy from anisotropy, being exposed simultaneously to chemical and thermal fluctuations of different intensities. In these cases, the enabling mechanism responsible for transduction of energy is typically the presence of a non-equilibrium steady state (NESS). A suitable stochastic model for such a phenomenon is the Brownian gyrator -- a two-degree of freedom stochastically driven system that exchanges energy and heat with the environment. In the context of such a model we present, from a stochastic control perspective, a geometric view of the energy harvesting mechanism  that entails a forced periodic trajectory of the system state on the thermodynamic manifold. Dissipation and work output are expressed accordingly as path integrals of a controlled process, and fundamental limitations on power and efficiency are expressed in geometric terms via a relationship to an isoperimetric problem. The theory is presented for high-order systems far from equilibrium and beyond the linear response regime.

\end{abstract}

\maketitle
\thispagestyle{empty}
\section{Introduction}
Carnot's 1824 work on heat engines \cite{carnot1824reflections} marked the birth of thermodynamics. Carnot hypothesized quasi-static reversible operation to derive limits on achievable efficiency, and led to the discovery of the concepts of entropy, absolute temperature, and irreversibility. In the two centuries since the publication of his pioneering work, in spite of great strides, attempts to analyze finite-time transitions were met with limited success. It was only recently when the emergence of Stochastic Thermodynamics \cite{seifert2012stochastic,sekimoto2010stochastic}, a subject at the interface between thermodynamics and stochastic control, allowed quantitative assessment of work and dissipation during fast thermodynamic transitions \cite{Jarz1996eq,Crooks1999FT}.

It was precisely this quantitative assessment that helped advance a geometric view of finite-time thermodynamics, in which dissipation is understood as a path-length traversed in a manifold of thermodynamic states. This view had its origins in macroscopic thermodynamics \cite{weinhold1975metric,Ruppeiner1995geom} and received renewed attention for linearized thermodynamic systems, in the so-called linear response regime~\cite{Crooks2007length,CrooksLengthControl2012,huang2020sub,Bradner2020geom,frim2021geometric}. However, it didn't reach its full potential until a relation between Stochastic Control, Thermodynamics and Optimal Mass Transport was drawn \cite{aurell2011optimal,aurell2012refined,chen2019stochastic,dechant2019thermodynamic}. This relation has enabled the expression of dissipation as the square of a Riemannian length, the 2-Wasserstein length, which has recently attracted a considerable amount of interest \cite{fu2020maximal,dechant2021geometric,taghvaei2021relation,dechant2022minimum,W2NakazatoEntropy2021,MarkovianW2Hasegawa2021,abiuso2022W2Carnotthermodynamics}.

Within this emerging stochastic control framework we herein present an analysis of the cyclic operation of a heat engine powered by anisotropy in its thermal environment. Specifically, we show that, for an $n$-dimensional overdamped system, dissipation and quasi-static work can be expressed as path integrals on the manifold of thermodynamic states. Our interest is in maximizing work output while driving the overdamped system over a cycle by periodic control. When the trajectory of our system lies on a two-dimensional manifold, the quasi-static work extracted over a cycle can be expressed as an area integral. Therefore, the problem of maximizing work output over a cycle amounts to maximizing an area integral subject to a fixed length of the closed thermodynamic path, which constitutes an isoperimetric problem. As a consequence, limits to efficiency are intrinsically related to an isoperimetric inequality.

We exemplify these results by revisiting our earlier work on the Brownian gyrator \cite{EnergyHarvestingAnisotropic2021}
--  an overdamped system in simultaneous contact with two heat baths along coupled degrees of freedom, where this circle of ideas was first brought out. Analogous isoperimetric results were derived in the linear response regime in \cite{thesisControlIsoperimLinear,ConferenceControlIsoperimLinear,frim2021geometric}, and in the context of quantum thermodynamics in \cite{alonso2022geometric}. We wish to emphasize that the results presented here apply to general systems outside the linearized regime.

\section{Preliminaries on optimal mass transport}\label{sec:OMT}
We outline certain geometrical notions from optimal mass transport~\cite{villani2003topics,ambrosio2005gradient} that play an essential role in the present paper. 

Given probability distributions $p_0$ and $p_f$ on $\mR^n$, 
\begin{equation}\nonumber
    \W_2(p_0,p_f)^2 := \inf_{\pi \in  \Pi(p_0,p_f)} \int_{\mR^n\times \mR^n} \|x-y\|^2  d \pi(x,y),
\end{equation}
where $\Pi(p_0,p_f)$ denotes the set of joint 
probability distributions on $\mR^n\times \mR^n$ with $p_0$, $p_f$ as marginals, defines the so-called $2$-Wasserstein metric. This induces a Riemannian metric, that we denote by $g_W$, on the tangent space of probability distributions on $\mR^n$ with finite second-order moments ($P_2(\mR^n)$), making $P_2(\mR^n)$ into a geodesic space. Geodesics correspond to (optimal) flows between endpoint distributions and $\W_2(p_0,p_f)$ becomes the geodesic length. 
Specifically, considering a time-varying probability distribution $p(t,x)$, driven by the velocity field $v(t,x)$ that together obey the continuity equation $\frac{\partial p}{\partial t}+\nabla_x \cdot(pv) = 0$, the integral
\begin{equation}\label{eq:action}
    \mathcal E[p,v]:= \int_0^{t_f}\int_{\mR^n} \|v(t,x)\|^2 p(t,x) \ud x \ud t,
\end{equation}
that may be seen as a cost on the control effort, in actuality represents action (i.e., ``kinetic energy'' integrated over time). A celebrated result by Benamou and Brenier \cite{benamou2000computational} links this cost functional to the Wasserstein metric, in that,
\begin{align}\label{eq:Benamou}
     &\min_{(p,v)}~\mathcal E[p,v] = \frac{1}{t_f}\W_2^2(p_0,p_f),
\end{align}
where the minimization is over pairs $(p,v)$ that satisfy the continuity equation and link $p_0$ to $p_f$.
The optimal velocity fields turn out to be curl-free and, hence, gradients $v=\nabla_x \phi$  (herein viewed as column vectors). These gradients can be identified with tangent vectors of $P_2(\mR^n)$ \cite[Sec.\ 8.1.2]{villani2003topics}.

\section{Stochastic thermodynamic model}
\label{sec:model}
We consider controlled $n$-dimensional overdamped  Langevin dynamics
\begin{equation}\label{eq:Langevin}
\gamma \ud X_t =  - \nabla_x U(t,X_t) \ud t + \sqrt{2\gamma k_BT}\ud B_t\quad X_0 \sim p_0,
\end{equation}  
with $X_t\in \mR^n$ representing location of particles, $p_0$ an initial distribution of an ensemble, and $k_B$ the Boltzmann constant. In addition, both  $\gamma$  and $T$ are diagonal matrices representing, respectively, the viscosity coefficient and the temperature of the ambient heat baths at which the different degrees of freedom are subjected. The terms in $\gamma$ will be assumed equal for simplicity, while the temperature $T$ must differ along directions in order to allow extracting energy -- it can be considered diagonal with distinct elements. Further, $B_t$ is an $n$ dimensional standard Brownian motion 
that models the thermal excitation from the heat baths, and $U(t,x)$ represents a time-varying potential exerting a force $-\partial_{x_i} U (t,x)$ along the $i$-th degree of freedom.
The potential function $U(t,x)$ that represents the control input, is externally controlled and helps exchange work with particles.

The probability density function  of $X_t$, denoted by $p(t,x)$,
constitutes the state of the system (thermodynamic ensemble) and satisfies the 
Fokker-Planck 
equation, 
\begin{align}\label{eq:FPK}
\frac{\partial p}{\partial t}  + \nabla_x \cdot\left(pv\right)=0,
\end{align}
where 
\begin{equation}
    v := -\gamma^{-1}(\nabla_x U + k_B T \nabla_x \log (p) + \chi), \label{eq:velocity-field} 
\end{equation}
and $\chi$ is any vector field such that $p\chi$ is divergence-free, that is $\nabla_x \cdot (p\chi)=0$. 
It turns out that $\chi$ can be chosen so that $v=\nabla_x \phi$, i.e., the gradient of a suitable function $\phi$ (see \cite[p.\ 247]{villani2003topics}), and thereby, $v$ is in the tangent space of $P_2(\mR^n)$. 
In fact, \[\phi=-\gamma^{-1}(U+\varphi), 
\] where $\varphi$ is the unique solution to $\nabla_x \cdot(p\nabla_x \varphi)= \nabla_x \cdot (pk_BT\nabla \log(p))$. 

The system~\eqref{eq:FPK} exchanges energy with the environment both through work done by changes in the external potential and through heat transfer with the thermal baths. The total energy of the system is $E = \int U(t,x) p(t,x) \,dx$, while the rate of work into the system due to  changes in the potential is
\begin{equation}\label{eq:work-p}
    \dot{W} = \int_{\mR^n} \frac{\partial U}{\partial t}(t,x) p(t,x)\, dx.
\end{equation}
The heat uptake from the respective thermal baths is
\begin{align*}
    \dot{Q}_i &= \int_{\mR}  \frac{\partial U}{\partial {x_i}}(t,x) \frac{\partial \phi}{\partial {x_i}}(t,x) \;p(t,x) dx_i
\end{align*}
resulting in the total heat uptake
\begin{equation}\label{eq:heat-p}
    \dot{Q} = \sum_{i=1}^n \dot{Q}_i=\int_{\mR^n}  (\nabla_x U)' \nabla_x\phi \, p dx,
\end{equation}
where $'$ denotes transpose.
Note that $\frac{d}{dt} E= \  \dot{W}+\dot{Q}$,
in agreement with the first law of thermodynamics.

One can use \eqref{eq:velocity-field} to write the gradient of the potential in terms of the state of the system, i.e.,
  $  \nabla_x U=-k_BT\nabla_x\log(p) -\gamma \nabla_x\phi-\chi
  $. Thus, the total heat uptake becomes
\begin{equation*}
     \dot{Q}=-k_B\hspace{-3pt}\int_{\mR^n}\hspace{-4pt}\big(\nabla_x\log(p)\big)'T'\nabla_x\phi\, p\,dx -\gamma\hspace{-3pt}\int_{\mR^n}\hspace{-4pt}\|\nabla_x\phi\|^2p\,dx,
\end{equation*}
where we have used the fact that the term $\int p\,\chi'\nabla_x\phi dx=0$. Integrating over the time interval $[0,t_f]$,
\begin{align*}
    Q=&-k_B\int_0^{t_f}\langle T\nabla_x(\log(p)),\nabla_x\phi\rangle dt\\&-\gamma\int_0^{t_f}\int_{\mR^n}\|\nabla_x\phi\|^2pdxdt,
\end{align*}
is expressed in terms of $p$ and $\nabla_x\phi$, with $\langle v_1,v_2\rangle:=\int_{\mR^n}v_1'v_2 p\,dx$ denoting inner product of vector fields.

Note that the integration in the last expression for $Q$ takes place over a time-indexed path $\{p(t,\cdot)\mid t\in[0,t_f]\}$ on the thermodynamic manifold. 
The first of the two integrals is linear in $\nabla_x\phi$ while the second is quadratic, converging to zero as the speed in traversing the path converges to zero. Therefore, the first term corresponds to the effective heat uptake in the quasi-static limit and the second corresponds to dissipation. 
Consequently, we define the quasi-static heat and dissipation over the interval $[0,t_f]$ as
  \begin{subequations}\label{eq:heat-general}  \begin{align}\label{eq:Q-qs}
  Q_{\rm qs}&:=-k_B\int_0^{t_f}\langle T\nabla_x(\log(p)),\nabla_x\phi\rangle dt,\\
    Q_{\rm diss}&:=\gamma\int_0^{t_f}\int_{\mR^n}\|\nabla_x\phi(t,x)\|^2p(t,x)\,dxdt,\label{eq:Q-diss}
\end{align}\end{subequations}
respectively, where both expressions are path integrals on $P_2(\mR^n),$ while dissipation can be identified with the value of the action integral~\eqref{eq:action} suitably scaled.

From the result by Benamou and Brenier \eqref{eq:Benamou} one can infer
    \begin{equation}\label{eq:min-diss}
\gamma\int_0^{t_f}\int_{\mR^n}\|\nabla_x\phi\|^2p\,dxdt \geq \frac{\gamma}{t_f} \W_2^2(p(0,\cdot),p(t_f,\cdot)). 
\end{equation} 
That is, the  minimum dissipation $Q_{\rm diss}$ is precisely the Wasserstein distance between the end-point distributions suitably scaled. 
The bound is achieved by transporting along the geodesic with constant velocity; this can be deduced from the Cauchy-Schwarz inequality
$$\gamma\int_0^{t_f}\!\!\!\int_{\mR^n}\!\|\nabla_x\phi\|^2p\,dxdt\geq \frac{\gamma}{t_f}\bigg(\!\int_0^{t_f}\!\!\!\sqrt{\int_{\mR^n}\|\nabla_x\phi\|^2p\,dx}\,dt\bigg)^2\!,$$
which holds as an equality when the velocity remains constant along a path.
The integral in parenthesis is the length of the curve $\{p(t,\cdot);t\in[0,t_f]\}$ in the Wasserstein-2 metric~\cite{aurell2011optimal,chen2019stochastic,fu2020maximal}, and is equal to the Wasserstein-2 distance when the curve is a geodesic.

\begin{remark}
The inequality $Q_{\rm diss}\geq 0$ encapsulates the second law of thermodynamics.
\end{remark}

\section{Geometry of cyclic processes}
\label{sec:cyclic}
Let us consider a cyclic transition of period $t_f$. From the first law of thermodynamics, 
the  work output is the difference between the quasi-static heat and the dissipative one, i.e.,
\begin{equation*}\label{eq:workout}
    W_{\rm out} = Q_{\rm qs} - Q_{\rm diss}.
\end{equation*}
Moreover, we define the efficiency of the cycle as the ratio between the work output and the maximum amount of work that can be extracted in a quasi-static setting \cite{Bradner2020geom}, this is,
\[
\eta = \frac{W_{\rm out}}{Q_{\rm qs}}.
\]
Note that as a consequence of the second law $\eta\leq 1$.
We can now cast the problem of maximizing work output over a cycle as the following stochastic control problem:
\begin{problem}\label{problem1} Determine
\begin{align*}
    &\max
    \{\, W_{\rm out}\ | \ \frac{\partial p}{\partial t}  +\nabla_x \cdot(p\nabla_x \phi)=0\,\}
\end{align*}
over all closed paths $\{p(t,\cdot)\mid t\in [0,t_f],\mbox{ with }p(0,\cdot)=p({t_f},\cdot)\}$ on the thermodynamic manifold $P_2(\mR^n)$.
\end{problem}

\vspace*{.1in}

Necessary conditions for optimality can be readily obtained by considering the Lagrangian
\begin{align*}
    \mathcal J=\int_0^{t_f}\!\!\!\int_{\mR^n} \!\!\!\Big(&k_B(T\nabla_x \log(p))'\nabla_x \phi \,p+\gamma \|\nabla_x \phi\|^2p\\
    &+\lambda(\frac{\partial p}{\partial t} +\nabla_x \cdot (p \nabla_x \phi))\Big) dx dt,
\end{align*}
with $\lambda(t,x)$ a Lagrange multiplier.
Its first variation is  \begin{align*}
   &\delta\mathcal J=\int_0^{t_f}\!\!\!\int_{\mathbb R^n} \!\!\!\Big(\!\!-\!\!\nabla_x \cdot\left(p\left[k_BT\nabla_x \log(p)\!+\!2\gamma \nabla_x \phi \!-\!\nabla_x\lambda \right]\right)\delta \phi\\&+\Big[\hspace{-1pt}-\hspace{-0.5pt}k_B\nabla_x\!\cdot\!(T'\nabla_x\phi)+\gamma\|\nabla_x \phi\|^2\hspace{-1.25pt}-\frac{\partial \lambda}{\partial t} \hspace{-0.5pt}-\hspace{-0.25pt}(\nabla_x \phi)'\nabla_x\lambda \Big]\delta p\\&+\Big[\frac{\partial p}{\partial t} +\nabla\cdot(p \nabla_x \phi)\Big]\delta \lambda\Big) dx dt,
\end{align*}
where we have used integration by parts. 
The first order necessary conditions for optimality are obtained by setting the variations to zero, namely
\begin{subequations}
\begin{align*}
    \nabla_x \phi&=\frac{\gamma^{-1}}{2}\left(\nabla_x\lambda- k_BT\nabla_x \log(p)-\chi\right),\\
    \frac{\partial \lambda}{\partial t} &=-(\nabla_x \phi)'\nabla_x\lambda+\gamma \|\nabla_x\phi\|^2-k_B\nabla_x\cdot(T'\nabla_x \phi),\\
    \frac{\partial p}{\partial t}  &= -\nabla_x \cdot(p\nabla_x \phi),
\end{align*}
\end{subequations}
where $\chi$ is a divergence-free vector-field as in~\eqref{eq:velocity-field}.

We consider Problem \ref{problem1} viewing trajectories as evolving on a two-dimensional submanifold $\mathcal M\subset P_2(\mR^n)$ with coordinates $(\lambda_1,\lambda_2)$; a similar two-dimensional problem has been previously studied in the linear response regime~\cite{frim2021geometric}. Thus,
we view $(\lambda_1,\lambda_2)$ as the two controlled degrees of freedom, while the state of our system
 $p(\lambda_1(t),\lambda_2(t),x)$,  $t\in[0,t_f]$, traverses a closed path on $\mathcal M$ encircling a domain $\mathcal D\subset \mathcal M$ and tracing its boundary $\partial \mathcal D$.

When $\partial \mathcal D$ is traversed with constant velocity by the curve $\{\alpha(t) = (\lambda_1(t),\lambda_2(t))\mid t\in[0,t_f]\}$, the quasi-static heat \eqref{eq:Q-qs} can be written as an area integral, and the dissipation \eqref{eq:Q-diss} expressed as a function of the length of the curve, i.e.
 \begin{align}\label{eq:QqsQdiss}
    Q_{\rm qs} = k_BT_r \Area_f,\qquad Q_{\rm diss} = \frac{\gamma\ell_r^2}{t_f} \ell^2,
\end{align}
where we have defined the weighted area of $\mathcal D$  and its perimeter by
\begin{align*}
    \Area_f &=\frac{1}{T_r}\oint_{\partial \mathcal D}-\bigg\langle T\nabla_x(\log(p)),\frac{\nabla_x\phi_t}{\hspace*{7pt}\|\nabla_x\phi\|_{L_{2,p}}} \bigg\rangle ds,\\
    \ell &= \frac{1}{\ell_r}\oint_{\partial \mathcal D} ds,
\end{align*}
where 
\[
\|\nabla_x\phi\|_{L_{2,p}}=\left(\int_{\mR^n}\|\nabla_x\phi\|^2p dx\right)^{1/2},
\]
and $ds=\|\nabla_x\phi\|_{L_{2,p}} dt$ represents the differential of the arc-length in $P_2(\mR^n)$ metrized with the Wasserstein $W_2$ metric;
$T_r$ and $\ell_r$ represent reference temperature and length, making the area and the length dimensionless quantities.

In terms of our two parameters, these two integrals can be written as

\begin{align*}
    \Area_f  &=\iint_{\mathcal D} f(\lambda_1,\lambda_2)\sqrt{\det(g_W)}d\lambda_1d\lambda_2,\\
    \ell&=\frac{1}{\ell_r}\oint_{\partial \mathcal D}ds,
\end{align*}
with the differential of the arc-length now expressed as $ds=\|\dot\alpha\|_{g_W}dt$, with
the 2-Wasserstein norm
\[
\|\dot{\alpha}\|_{g_W}:= ( \dot{\alpha}'g_W\dot{\alpha})^{1/2}
\]
    of the velocity of the curve $\alpha$ expressed with a quadratic form using the matrix $g_W$. Thence, $f(\lambda_1,\lambda_2)$ is a work density relative to the canonical $2$-form $\sqrt{\det(g_W)}d\lambda_1 d\lambda_2$, that can be obtained using Stokes' theorem.

Consequently, we can write the work output $W_{\rm out}$ and efficiency $\eta$ in terms of the area and length as
\begin{equation}
     W_{\rm out}=k_BT_r\big(\Area_f-\mu \ell^2\big)\quad\mbox{and}\quad \eta = 1-\mu\frac{\ell^2}{\Area_f},
     \label{eq:defefficiency}
\end{equation}
where
 $\mu=\frac{t_c}{t_f}$ is a dimensionless constant, with  $t_c = \frac{\gamma \ell_r^2}{k_B T_r}$ the {characteristic time} that a Brownian motion with intensity $\sqrt{\gamma^{-1}k_BT_r}$ needs to traverse a distance $\ell_r$ on average.  Thus, the problem of maximizing work output over a cycle on the manifold of thermodynamic states can be recast as
\begin{equation}\label{eq:maxwork}
    W^*(\mu):=k_BT_r\max_{\mathcal D} ~\{\Area_f- \mu \ell^2\},
\end{equation}
for different values of $\mu$. Maximization of $\Area_f-\mu\ell^2$ relates to the isoperimetric problem, which is that of maximizing area for a fixed length of its perimeter, this is,
\begin{problem}\label{problem2}
Determine
\begin{equation*}\label{eq:maxarea}
    \Area_f^*(\ell) :=\max\{\Area_f \mid   \ell \mbox{ is specified}\},
\end{equation*}
over all closed paths encircling a domain $\mathcal D\subset \mathcal M$.
\end{problem}
One can understand the relationship between \eqref{eq:maxwork} and the isoperimetric problem \ref{problem2} by viewing
$\mu$ in \eqref{eq:maxwork} as a Lagrange multiplier for Problem \ref{problem2}. 

We obtain a first-order condition for the isoperimetric problem \ref{problem2} that characterizes optimal cycles  through variational analysis. 
To this end, we parametrize the closed curve $\alpha(\cdot)$ tracing $\partial \mathcal D$ by the arclength $s$.
We let $ds$ and $du$ denote the differentials along the curve and normal to the curve respectively, so that the corresponding local coordinates form an orthonormal system.
Under a perturbation $\alpha(s)\to \alpha(s) + \psi(s)\hat n(s)du$, where $\hat n(s)$ is the (outward) normal unit vector at $s$ and $\psi(\cdot)$ is an arbitrary scalar function, the perimeter is perturbed to $\int_{s=0}^\ell (1+\kappa(s) \psi(s)du)ds$, where $\kappa(\cdot)$ denotes the geodesic curvature~\cite{morgan1998riemannian}.

%
%
Hence, the variation of $\ell^2$ is
$
\delta \ell^2 = 2\ell \int_{s=0}^\ell \kappa(s)\psi(s) dsdu
$. 
On the other hand, as the domain $\mathcal D$ is enlarged to $\mathcal D\cup\delta \mathcal D$,
\begin{align*}
\delta \Area_f &= \iint_{\delta\mathcal D}f(\lambda_1,\lambda_2)\sqrt{\det(g_W)}d\lambda_1 d\lambda_2\\
&= \int_{s=0}^\ell  
f(\lambda_1(s),\lambda_2(s))\psi(s)dsdu.
\end{align*}
Thus, the first-order optimally condition $\delta \mathcal A_f-\mu\delta \ell^2=0$ gives that the ratio of the geodesic curvature $\kappa$ over the density 
$f$
must be constant and equal to $1/(2\ell \mu)$ at each point of the curve that traces $\partial \mathcal D$, this is, \begin{equation}\label{FONC}
    \frac{\kappa(\lambda_1,\lambda_2)}{f(\lambda_1,\lambda_2)}=\frac{1}{2\ell\mu}.
\end{equation}

The solution to the isoperimetric problem, $\Area_f^*(\ell)$, helps answer a variety of questions on optimizing control strategies. 
Specifically, the maximal work output
\begin{equation*}
    W^*(\mu)=k_BT_r\max_{\ell}\{\mathcal A_f^*(\ell)-\mu \ell^2\}
\end{equation*} 
takes place where $d\Area_f^*(\ell)/d\ell^2=\mu$.
Also, $\Area_f^*(\ell)$ allows computing the maximal work for a given efficiency $\eta$
\begin{equation*}
    W^*(\mu)|_\eta=k_BT_r\eta \max_\ell\{\Area_f^*(\ell)\ | \ \eta=1-\mu\frac{\ell^2}{\Area_f^*(\ell)}\},
\end{equation*}
which is obtained for $\ell$ such that $\mathcal A_f^*(\ell)=\frac{\mu}{1-\eta}\ell^2$. The existence of these operating points is guaranteed if $\lim_{\ell \to \infty}\Area_f/\ell^2 = 0$, which is satisfied for manifolds with strictly negative curvature, as we will see in the following.


%
 \begin{remark}
 Even if the relationship to the isoperimetric problem is particular to two dimensions, it should be pointed out that, given \eqref{eq:Q-qs} and \eqref{eq:Q-diss},  as long as our state manifold is finite dimensional, one can always use Stokes' theorem and write the work output and the efficiency in terms of a trade-off between area and length.
 \end{remark}

Let us now take a look back at the expression for efficiency in terms of area and length~\eqref{eq:defefficiency}. The problem of bounding efficiency is equivalent to that of finding an isoperimetric inequality in the space of thermodynamic states.

Isoperimetric inequalities, for a simply-connected domain $D$ with area $A$ and length $l$ on a manifold with Gaussian curvature $G$, take the form~\cite{osserman1978isoperimetric}
 \begin{align*}
 l^2&\geq 4\pi A- 2\left(\iint_DG^+\right)A,\\
     l^2&\geq 4\pi A- (\sup_{\mathcal D} G )A^2,
 \end{align*} 
 among others, where $G^+(p)=\max \{G(p),0\}$. We illustrate how isoperimetric inequalities can be used to bound efficiency by specializing to the case when our manifold of thermodynamic states has negative curvature. Indeed, if $G\leq 0$, then
$
     \ell^2\geq 4\pi \mathcal A,
$
 where $\mathcal A=\iint_{\mathcal D} \sqrt{\det(g_W)} d\lambda_1\lambda_2$ is the area of $\mathcal D$ with respect to the canonical $2$-form of $\mathcal M$. Thus, we obtain the following result:
 \begin{proposition}
      Consider a two-dimensional submanifold $\mathcal M\subset P_2(\mR^n)$ with coordinates $(\lambda_1,\lambda_2)$ that has negative Gaussian curvature. Then, efficiency of a thermodynamic cycle on $\mathcal M$ can be bounded as follows:
 \begin{equation*}\label{eq:isoperimetric}
     \eta=1-\mu\frac{\ell^2}{\Area_f}= 1-\mu\frac{\ell^2 }{\Area }\frac{\Area}{\Area_f}\leq 1-\frac{4\pi} {\bar f }\frac{t_c}{t_f},
 \end{equation*}
 where $\bar f=\Area_f/\Area$.
 \end{proposition}
 \begin{remark}
 A conservative estimate for $\bar f$ is $\max_{\lambda_1,\lambda_2}f$. The above bound depends on physical parameters as well as on the chosen period, and turns negative when positive work output is not possible. In particular, the bound implies that if $t_f\leq 4\pi t_c/\bar f$, it is impossible to extract positive work. Thus, $4\pi t_c/\bar f$ {\em constitutes a threshold for the period of work-producing cycles}.
 \end{remark}
 
  On the other hand, if the Gaussian curvature of our manifold $\mathcal M$ is strictly negative, that is, $G\leq -k$ for some $k>0$, then we can conclude a tighter bound, namely,
$
      \ell^2\geq 4\pi \Area+k\Area^2,
$
or, in terms of $\Area_f$,
 \begin{equation*}
      \ell^2\geq \frac{4\pi}{\bar f} \Area_f+\frac{k}{\bar f^2}\Area_f^2.
 \end{equation*}
 Specifically, this implies that the limit $\lim_{\ell \to \infty}\Area_f/\ell^2 = 0$, and thus, the optimal operating points for the problem of maximizing work output and maximizing work output with a given efficiency exist. An example of a thermodynamic manifold with negative curvature appears in \cite{frim2021geometric}, in the linear response regime.

\section{The example of a Brownian Gyrator}

We now exemplify the usefulness of the above results by revisiting the following two-dimensional overdamped system, which was discussed in our previous work \cite{EnergyHarvestingAnisotropic2021}:
\begin{equation}\tag{3'}
    \begin{aligned}
	d x_t &= - \gamma^{-1}\nablax U(t,x,y) d t + \sqrt{{2\gamma^{-1}k_BT_x}} d B^x_t,
\\
		d y_t &= - \gamma^{-1} \nablay U(t,x,y)d t + \sqrt{{2\gamma^{-1}k_BT_y}} d B^y_t,
\end{aligned}
\end{equation}
where $\{B^x_t\}_{t\geq 0}$ and $\{B^y_t\}_{t\geq 0}$ are two independent standard Brownian motions, while $T_x$ and $T_y$ represent temperature along each of the two degrees of freedom $x$ and $y$, respectively. Note that the usage of prime in labeling equations highlights correspondence with equations in the earlier sections.
The potential $U(t,x,y)$ is assumed to be quadratic
\begin{equation*}
    U(t,x,y)=\frac{1}{2} \xi' K(t)\xi, \quad \mbox{where}\quad \xi=\left[\begin{array}{c}x\\ y\end{array}\right],
\end{equation*}
with $K(t)$ a symmetric $2\times 2$ matrix seen as a control variable. Without loss of generality, we define $T_r :=(T_x-T_y)/2>0$.

If the initial state is Gaussian ${\mathcal N}(0,\Sigma_0)$, i.e., $0$-mean with covariance $\Sigma_0$, then it remains Gaussian with $0$-mean and convariance $\Sigma(t)$ that satisfies the Lyapunov equation
\begin{equation}\label{eq:Lyapunov}
    \gamma \dot\Sigma(t)=-K(t)\Sigma(t)-\Sigma(t) K(t)+2k_BT,\tag{4'}
\end{equation}
which is derived from the Fokker-Planck eq.~\eqref{eq:FPK}, where
$$T:=\left[\begin{array}{cc}T_x&0\\ 0&T_y\end{array}\right]. $$
Thus, the state of the system $p(t,x)$ can be identified with the covariance matrix $\Sigma(t)$, a symmetric $2\times 2$ matrix.

In order to determine the velocity field $\nabla_x\phi$, we use the ansatz that $\phi(t,\xi)=\frac12 \xi^T \Phi(t) \xi$, with $\Phi(t)$ symmetric and that $\chi(t,\xi)=A(t)\xi$. The condition $\nabla \cdot (p\chi)=0$ translates to $A(t)\Sigma(t) =:J(t)$ being skew symmetric. Then, \eqref{eq:velocity-field} can be written as
\begin{equation}\tag{5'}
    \Phi(t)  = -\gamma^{-1}\left( K(t) + k_BT\Sigma(t)^{-1} + J(t)\Sigma(t)^{-1}\right),
\end{equation}
where the skew-symmetric matrix $J(t)$ is selected so that $\Phi(t)$ becomes symmetric.
Using the Lyapunov equation~\eqref{eq:Lyapunov} we deduce that $\Phi(t)$ solves $\dot\Sigma(t)=\Sigma(t)\Phi(t)+\Phi(t)\Sigma(t)$, and thus, 
$$
\Phi(t)=\mathcal L_{\Sigma(t)}[\dot\Sigma(t)]:=\int_0^\infty   e^{-\tau \Sigma(t)} \dot{\Sigma}(t)e^{-\tau \Sigma(t)}d\tau.
$$
Given that $\nabla_x\phi(t,\xi)=\mathcal L_{\Sigma(t)}[\dot\Sigma(t)]\xi$, the expressions for the quasi-static and dissipative heat in \eqref{eq:heat-general} translate to
\begin{align}\tag{8a'}\label{eq:Q-qs-polar}
  \mathcal Q_{\rm qs}&
  =k_B\int_0^{t_f}\trace[T\mathcal L_\Sigma[\dot\Sigma]]dt,\\ \tag{8b'}
  \mathcal Q_{\rm diss}&
  =\gamma\int_0^{t_f}\trace[
  \Sigma(\mathcal L_\Sigma[\dot\Sigma])^2]dt.\label{eq:Q-diss-polar}
\end{align}

In order to utilize the results in Section \ref{sec:cyclic}, we restrict the controlled degrees of freedom on the state manifold (identified with the $\Sigma$-space) to $2$, by imposing that $\det(\Sigma(t))$ be constant. Under this restriction, the $2\times 2$ positive definite covariance can be written in polar coordinates $(\lambda_1=r,\lambda_2=\theta)\in [0,\infty)\times [0,2\pi)$ as
\begin{equation*}\label{eq:Sigmartheta}
    \Sigma(r,\theta)=R\Big(\hspace{-3pt}-\frac{\theta}{2}\Big)\sigma^2(r)R\Big(\frac{\theta}{2}\Big),
\end{equation*}
where $R(\cdot)$ and $\sigma^2(\cdot)$ are orthogonal and diagonal matrices, respectively, given by
\begin{equation*}
    R(\vartheta)\hspace{-1pt}=\hspace{-1pt}\left[\begin{array}{cc}  \hspace{-5pt}\cos(\vartheta) \hspace{-4pt}&  \hspace{-2pt} \sin(\vartheta)\hspace{-4pt}\\  \hspace{-5pt}-\sin(\vartheta) \hspace{-4pt}& \hspace{-2pt}\cos(\vartheta)  \hspace{-4pt}\end{array}\right]~\mbox{and}~~
   \sigma^2(r) \hspace{-1pt}=\hspace{-1pt}\left[\begin{array}{cc}  \hspace{-5pt}2\lam^2e^r\hspace{-4pt}&  \hspace{-4pt}0\hspace{-4pt}\\
   \hspace{-4pt}0\hspace{-4pt}&  \hspace{-4pt}2\lam^2e^{-r}\hspace{-6pt}\end{array}\right],
\end{equation*}
where $\lam=\sqrt[4]{\det(\Sigma(t))/4}$\label{page:charlength} 
represents a
(constant)  characteristic length for the system. 

We consider the integrals (\ref{eq:Q-qs-polar}-\ref{eq:Q-diss-polar}) over a cycle that encircles a domain $\mathcal D$, over the boundary $\partial \mathcal D$ that is traced by the curve $\{\alpha(t) = (r(t),\theta(t))\mid t\in[0,t_f]\}$ with constant velocity. Then, the quasi-static heat and dissipation can be written in terms of the parametrization as in \eqref{eq:QqsQdiss}, with
 \begin{equation*}
      g_W \hspace{-2pt}= \hspace{-2pt}\left[\begin{array}{cc}
           \hspace{-2pt}\cosh(r) \hspace{-2pt}& \hspace{-2pt}0 \hspace{-2pt}\\
         \hspace{-2pt} 0 \hspace{-2pt}& \hspace{-2pt}\frac{\sinh^2(r) }{\cosh(r)} \hspace{-2pt}
     \end{array}\right]\ \mbox{and}\ f(r,\theta)=\frac{\sin(\theta)\sinh(r)}{\cosh^2(r)},
 \end{equation*} 
where $f$ has been obtained through Stokes' theorem (see \cite{EnergyHarvestingAnisotropic2021} for the detailed computation). Similarly, the work output and efficiency are given by \eqref{eq:defefficiency}, and the problem of maximizing work output is related to  the isoperimetric problem \ref{problem2}.

 \begin{figure}[t]
     \centering
     \includegraphics[width=0.45\textwidth,trim={0cm 0cm 0cm 0cm},clip]{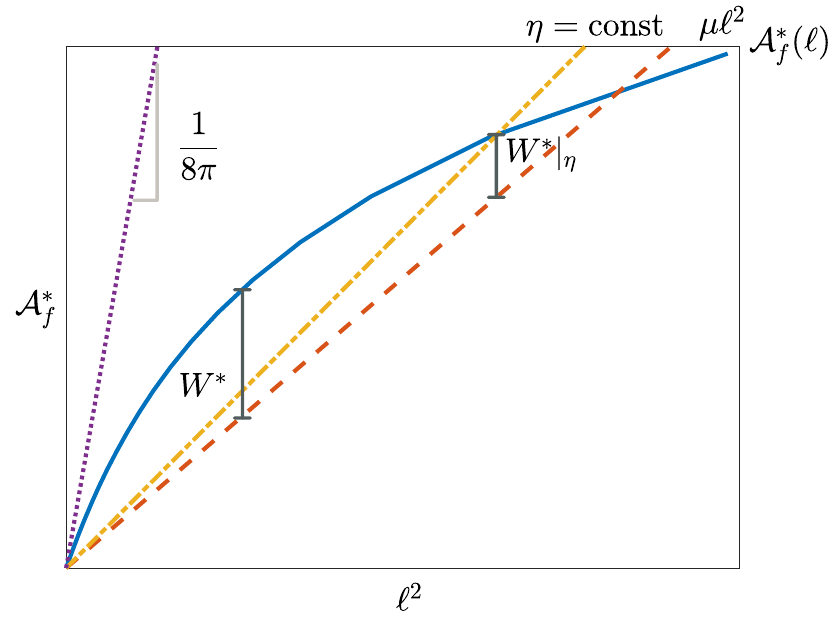}
     \caption{Maximum area $\mathcal A_f^*(\ell)$ after solving the isoperimetric problem \ref{problem2} for the Brownian gyrator case, shown with solid blue curve. The maximal work output $W^*(\mu)$ 
corresponds to the maximal vertical distance between  $\mathcal A_f^*(\ell)$ and the line $\mu \ell^2$, which takes place where $d\mathcal A_f^*(\ell)/d\ell^2=\mu$.
On the other hand, operating points with efficiency $\eta$ provide work $W_{\rm out}=k_B T_r\eta \mathcal A_f$ and lie on the line $\mathcal A_f=\frac{\mu}{1-\eta}\ell^2$ shown (dash-dotted). Therefore, the intersection of this line with the (blue) curve $\mathcal A_f^*(\ell)$ gives the sought optimal operating point for a given efficiency.  }
     \label{fig:A-lsq}
 \end{figure}

\begin{figure}[t]
     \centering
     \includegraphics[width=0.45\textwidth, trim={1.5cm 21.5cm 12cm 2cm}, clip]{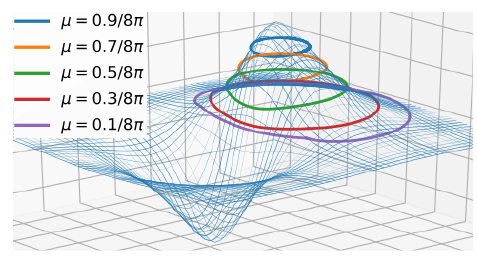}
     \caption{ The figure depicts the $f$-surface in polar coordinates  $(r,\theta)$ (drawn as a blue grid). Optimal cycles of thermodynamic states solving the isoperimetric problem for different values of $\mu=2\gamma \ell_r^2/(k_BT_rt_f)$ are depicted with closed curves on the $f$-surface. 
    }
     \label{fig:max-work-amir}
 \end{figure}

 Figure \ref{fig:A-lsq} shows the relationship between $\mathcal A^*_f$ and the operating points at which both the maximum work output $W^*$ and the maximum work output for a fixed efficiency $W^*|_\eta$ are obtained. Indeed, in this setting, the existence of these points can be ensured \cite{EnergyHarvestingAnisotropic2021}. On the other hand, Figure~\ref{fig:max-work-amir} displays several solutions to the isoperimetric problem that have been obtained numerically using the first-order optimality condition~\eqref{FONC}.  It is observed that as $\mu$ becomes small, and thus, the corresponding penalty on the length decreases,
the area that the optimal cycle encircles increases. In contrast, as $\mu$ becomes large, the optimal cycle shrinks to the point, beyond which (i.e., for larger $\mu$) it is impossible to extract positive work, which points to an isoperimetric inequality.
 
Indeed, even if in this case the Gaussian curvature of the Riemannian manifold $\mathcal M$ \cite[page 23]{morgan1998riemannian},\cite[page 400]{Gray2006diffgeom},
$\nonumber
G(r,\theta)=1/\cosh^3(r),
$
is nonnegative, we can still use this framework to bound efficiency using the following isoperimetric inequality for manifolds with rotationally symmetric metric
\cite[Page 113]{morgan1998riemannian}:
\begin{equation*}\label{eq:iso}
\ell^2 \hspace*{-2pt}\geq 4\pi{\Area} -2\int_0^{\Area}\bar G(\tau)d\tau,
\end{equation*}
where $\bar G(\tau)$ is the area integral of the Gaussian curvature over a circle centered at the origin with area $\tau$. 
Then, \begin{align*}
\ell^2&\geq 4\pi\Area- 4\pi^2\bigg(\frac{\Area}{2\pi}-\log\Big(1+\frac{\Area}{2\pi}\Big)\bigg)
\geq 2\pi \Area=\frac{2\pi}{\bar f} \Area_f.
\end{align*}
Since  $   \Area_f\leq  \max_{(r,\theta)\in\mathcal D} ~f(r,\theta) \times \Area=\frac12 \Area
$, we obtain
\begin{equation*}
    \eta \leq 1-4\pi\mu.
\end{equation*}
This result was obtained in \cite{EnergyHarvestingAnisotropic2021} where the tighter bound $\eta\leq 1-8\pi\mu$ was also conjectured.\\

\section{Concluding Remarks}

Geometric frameworks for understanding thermodynamics of macroscopic systems are classical and well established.
Recent insights in modeling microscopic thermodynamic systems via stochastic differential equations (Langevin dynamics), through Stochastic Thermodynamics, have enabled an analogous geometric view of microscopic systems as well. Early contributions focused mostly on thermodynamic transitions near equilibrium. Herein we study, beyond the linearized regime, an $n$-degree of freedom (DoF) system in simultaneous contact with  heat baths of different temperature. The anisotropy in thermal fluctuations, in conjunction with the coupled DoF, allow energy extraction from the heat baths. 
Our contribution in this paper lies in pointing to geometric interpretations for power and efficiency for finite-time thermodynamics, paving the way for new insights in the design of heat engines and, perhaps, for interpreting the functionality of molecular engines in the biological world that are powered by anisotropy in chemical potentials.\\

{\em Acknowledgments.}\hspace*{5pt}
Partial support by ”la Caixa” Foundation (ID 100010434) with code LCF/BQ/AA20/11820047, NSF under grants
1839441, 1901599, 1942523, and AFOSR under grant FA9550-20-1-0029 are gratefully acknowledged.

\bibliographystyle{IEEEtran}
\bibliography{mainv2}



\end{document}